\pdfoutput=1
\documentclass[%
 aip,
 jmp,%
 amsmath,amssymb,
 preprint,%
 reprint,%
]{revtex4-1}
\usepackage{CJK}
\usepackage{graphicx}
\usepackage{epstopdf}
\usepackage{dcolumn}
\usepackage{bm}
\CJKnospace

\begin{document}

\title[]{An Experiment About Parallel Circuit And The Lorentz Forces On Wires}

\begin{CJK}{UTF8}{song}
\thanks{The following article has been submitted to The Physics Teacher. After it is published, it will be found at \url{http://scitation.aip.org/content/aapt/journal/tpt}.}
\CJKfamily{gbsn}
\author{Audrey Yueru Li(李乐如)}
\affiliation{Student at Cabin John Middle School, Potomac, Maryland, USA}
\author{Shengchao Alfred Li(李盛超)}
\email{Shengchao.Li@gmail.com}
\affiliation{Amateur scientist in Potomac, Maryland, USA}
\begin{abstract}
Parallel circuit \cite{:/content/aapt/journal/tpt/41/2/10.1119/1.1542048, :/content/aapt/journal/tpt/43/7/10.1119/1.2060644} and the Lorentz forces on current carrying wires \cite{:/content/aapt/journal/tpt/38/5/10.1119/1.880536, :/content/aapt/journal/tpt/45/5/10.1119/1.2731270} are important concepts in introductory physics courses. Here we describe an experiment that illustrates these two concepts. We mount a circuit with multiple grounding points onto a torsion balance. We show that the grounding points create parallel return paths for the supply current. When the topology or the shapes of the return paths are altered, the Lorentz forces exerted by the currents in the return paths within a magnetic field change accordingly, which in turn cause changes in the rotary displacement of the torsion balance. This experiment is simple and can be easily reproduced in a teaching laboratory. What makes it interesting to students is that recently two research teams have attempted to detect thrusts from microwave driven asymmetrical resonance cavities (``EmDrive" or ``Cannae Drive")\cite{Brady2014, Tajmar2015}, and the phenomenon observable in this experiment provides an alternative explanation to the ``thrusts" they detected.
\end{abstract}
\keywords{EmDrive, Cannae Drive, Lorentz force, parallel circuit, ground loop, torsion balance}
\maketitle
\clearpage\end{CJK}


\section{The Proof-of-Concept Experiment Apparatus\cite{SM,SMzip}}

\begin{figure}
\includegraphics[scale=0.83]{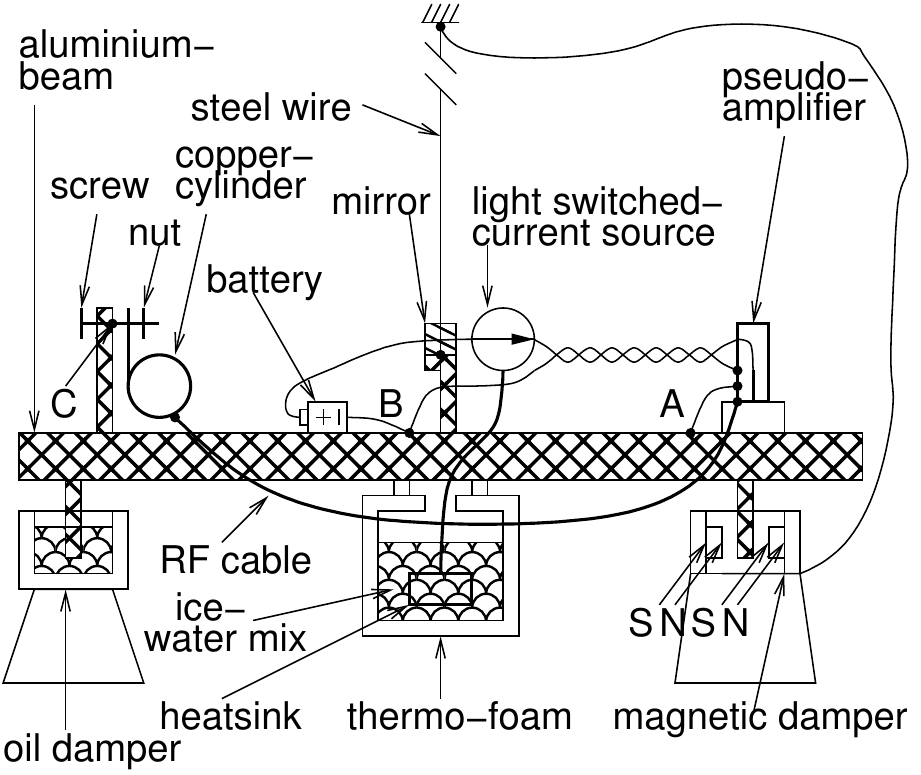}
\caption{\label{fig:circuitNew}The proof-of-concept experiment apparatus.}
\end{figure}

Our initial design of the experiment apparatus (FIG.~\ref{fig:circuitNew}) is for the proof of concept. We show that it is capable of detecting small Lorentz force differences when grounding points of a circuit are altered. In this design, we pay special attention to avoiding potential disturbances. All parts except for a hanging steel wire are either non-ferromagnetic or only very weakly ferromagnetic. Air and thermal disturbances are also controlled for. We simplify the experiment apparatus in section \ref{sec:Simple}. 

This design is explained as follows. We make a torsion balance by attaching the center of an aluminium beam to a hanging steel wire, similar to what Cavendish once did \cite{:/content/aapt/journal/tpt/27/7/10.1119/1.2342871}. We attach a mirror at the lower end of the steel wire. A laser beam deflected by the mirror magnifies the torsion balance's small horizontal rotary displacement around its center. A magnetic damper is installed at one end of the beam, consisting of magnets forming a gap and magnetic field within and beyond the gap, and an aluminium fin with one end attached to the beam and the other end plugged into the gap. The magnetic damper works because the electrical current (the ``Eddy" current) incurred in the aluminium fin always impairs its movement (Faraday's Law). An oil damper is installed at the other end of the beam, consisting of a cup of oil and another aluminium fin with one end attached to the beam and the other end plugged into the oil. A light switched current source of 5.60 Amperes and a pair of twisted power lines allow electrical current to flow from a battery pack to a pseudo amplifier, which consists a copper shell and nothing else. An RF cable connects with its shield layer the pseudo amplifier and a copper cylinder with end caps. A heat sink helps to dissipate heat generated by the current source. The heat sink is submerged in a thermally insulated container filled with ice and water mix, so the temperature of the heat sink is kept constant during the courses of tests, to avoid potential thermal disturbance. The container is stuffed with coarse sponge to discourage relative rotary movement between the ice and water mix and the container, while allowing heat to disperse within the mix. The whole experiment apparatus is enclosed in a chamber to avoid air flow disturbance.

\begin{figure}
\includegraphics[scale=0.83]{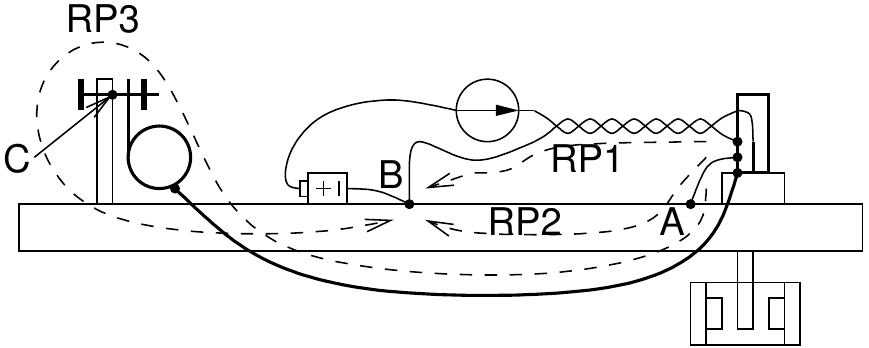}
\caption{\label{fig:circuitNewPaths} The return paths. Dashed lines indicate the return paths. RP1, return path 1. RP2, return path 2. RP3, return path 3. Return path 2 and 3 exist because of three grounding points, point A, point B and point C.}
\end{figure}

\begin{figure}
\includegraphics[scale=0.83]{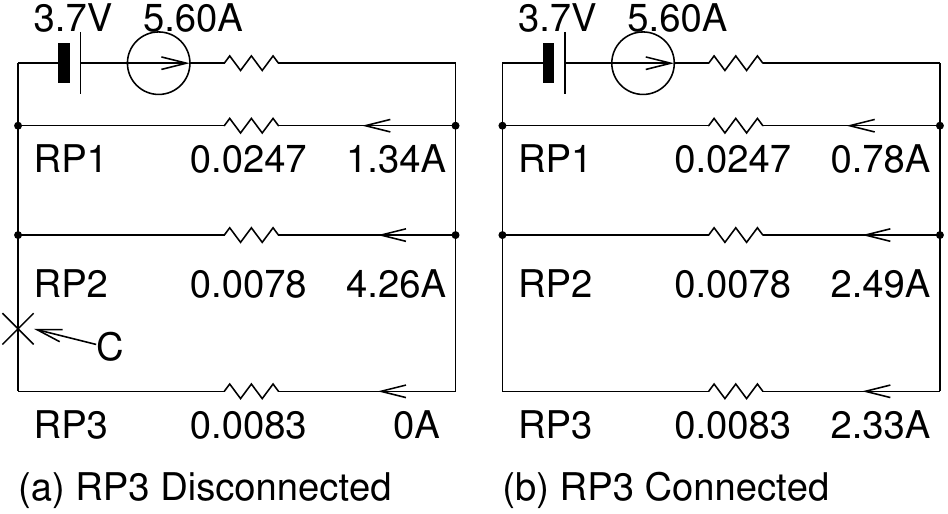}
\caption{\label{fig:circuitCurrent1} Currents in the return paths. Resistance of each return path is measured by feeding 5.60 A current into that path. Currents in the return paths are calculated. RP1, return path 1. RP2, return path 2. RP3, return path 3. Resistance unit is $\Omega$. (a), return path 3 is disconnected. (b), return path 3 is connected. }
\end{figure}

There are three grounding points, point A, point B and point C. Three return paths exist because of these grounding points (FIG.~\ref{fig:circuitNewPaths}). Return path 1 starts from the shell of the pseudo amplifier, goes through one of the twisted power lines, point B, and finally goes back to the battery pack. Return path 2 starts from the shell of the pseudo amplifier, goes through point A, the portion of the beam between point A and point B, point B, and finally goes back to the battery pack. Return path 3 starts from the shell of the pseudo amplifier, goes through the shield layer of the RF cable, the copper cylinder, point C, the portion of the beam between point C and point B, point B, and finally goes back to the battery pack. We measure the resistances of the return paths and calculate currents in the paths in two situations. One is with point C electrically cut open thus return path 3 disconnected, the other is with point C electrically connected thus return path 3 connected. The results are shown in FIG.~\ref{fig:circuitCurrent1}. 

Because of the existence of the Earth's magnetic field and the magnetic damper's magnetic field, The currents in the return paths exert the Lorentz forces, which in turn cause a torque on the torsion balance, which can be measured from the rotary displacement of the torsion balance. We define the counter-clock rotation as positive when we look down at the torsion balance from above.

In this paper, all orientations are relative to the Earth's magnetic field.

\section{The proof-of-concept test results\cite{SM,SMzip}}

We carried out three tests with the proof-of-concept experiment apparatus.

\begin{figure}
\includegraphics[scale=0.75]{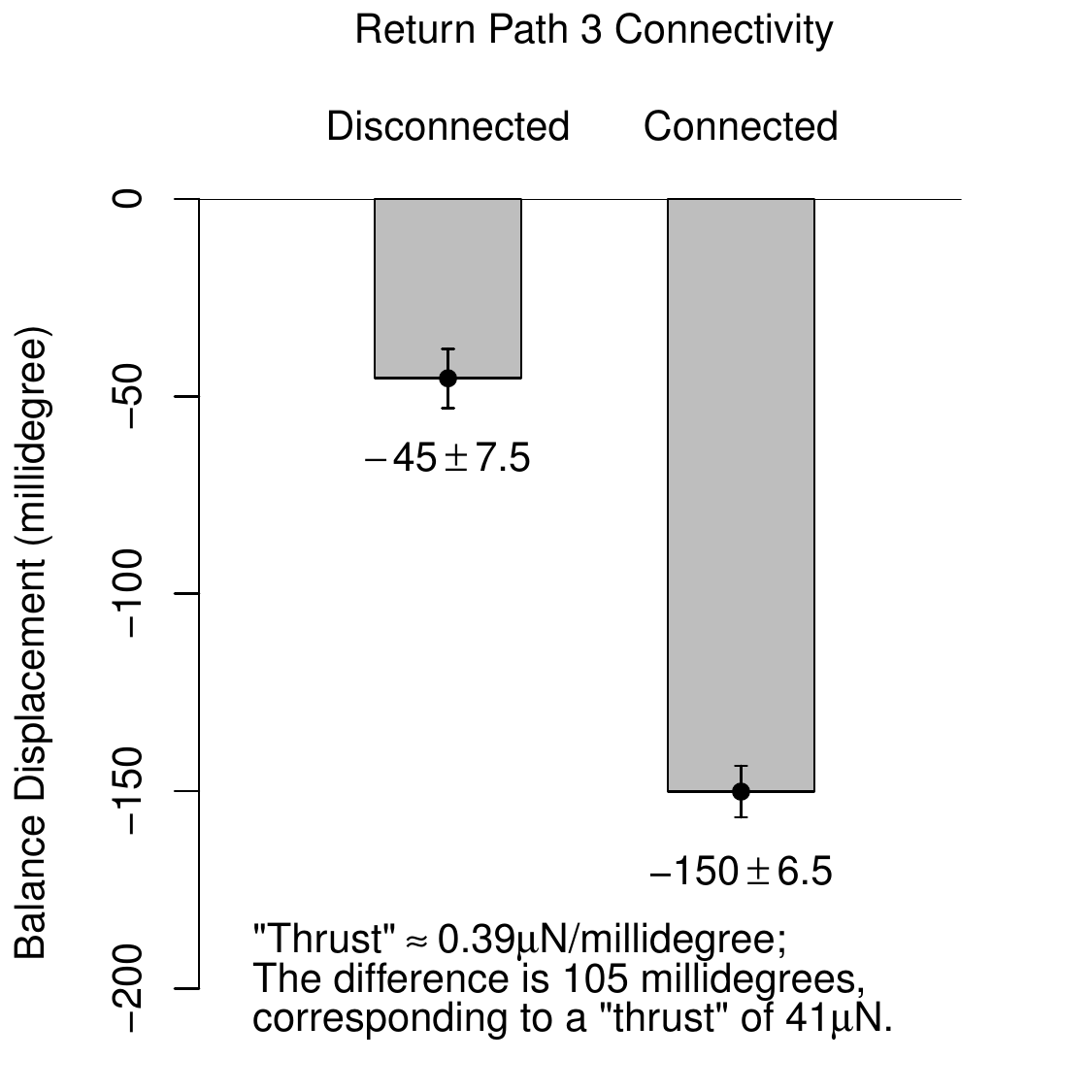}
\caption{\label{fig:TestA}Test A results. The torsion balance pointed to the North. There were 5 data points of each bar. ``Thrust" calibration was taken from the average of Test B and Test C.}
\end{figure}

In Test A, We aligned the torsion balance so its point A end pointed to the North. The damper magnets were in the orientation as shown in FIG.~\ref{fig:circuitNew}. We changed return path 3 connectivity at point C, alternatively between being connected and being disconnected, 10 times in total, 5 times for each. Each time we measured the torsion balance's rotary displacement when the current source was switched on. We grouped the measured rotary displacements according to the connectivity of return path 3, and made inference on whether the torsion balance's rotary displacements were the same in the two situations. A two sided t-test gave a p-value of $1.4\times 10^{-8}$, with which we rejected the idea that they were the same. The results are illustrated in FIG.~\ref{fig:TestA}.

\begin{figure}
\includegraphics[scale=0.75]{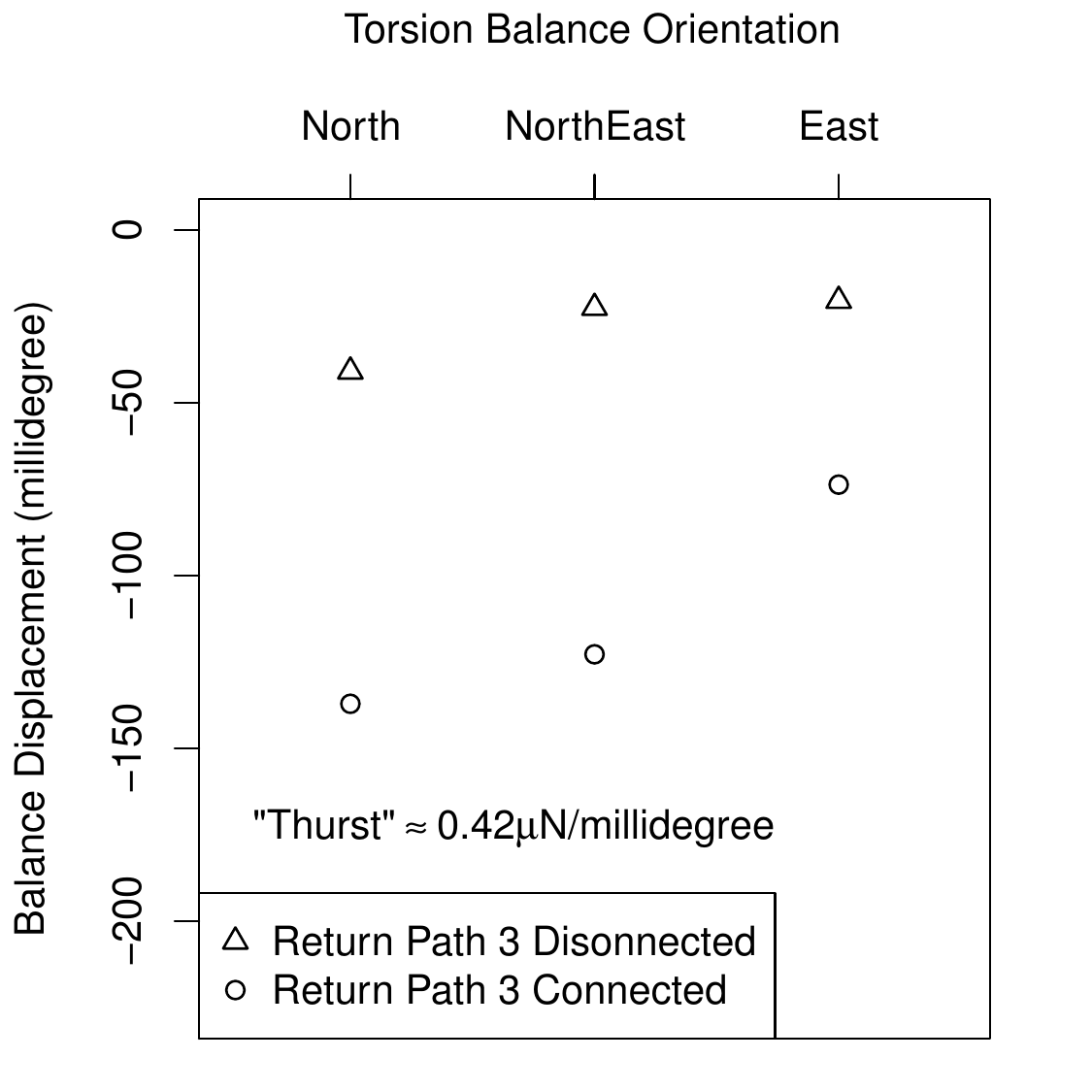}
\caption{\label{fig:TestB} Test B results. The influence of the Earth's magnetic field was evident. The Northeast is 50 degrees from the North.}
\end{figure}

In Test B, we aligned the torsion balance to the North, to the East and to the Northeast, alternatively. Each time we measured the torsion balance's rotary displacements with return path 3 disconnected and connected, respectively. The results are illustrated in FIG.~\ref{fig:TestB}. The results suggested that the Earth's magnetic field had the potential to influence the torsion balance's rotary displacement when the current source was switched on.

\begin{figure}
\includegraphics[scale=0.75]{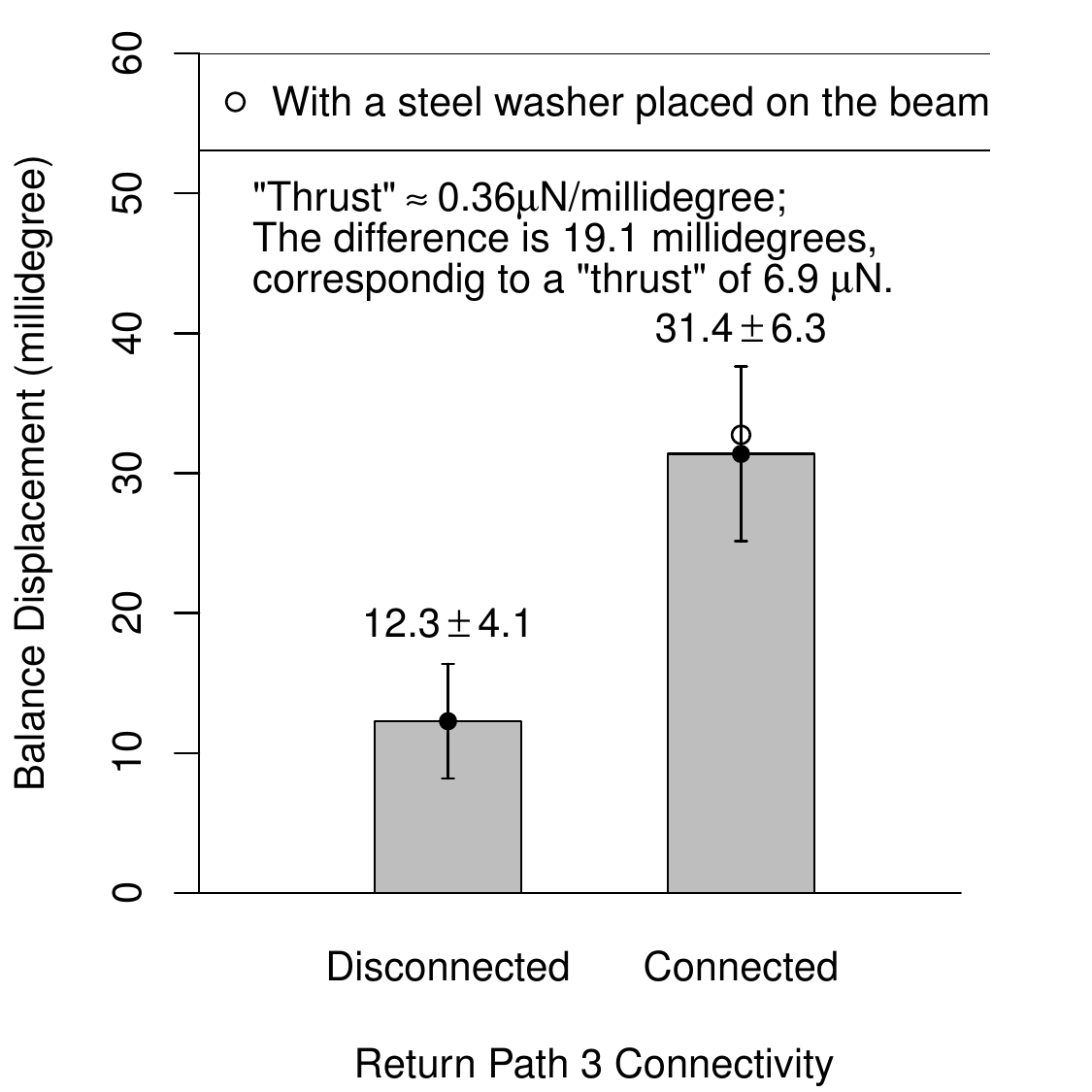}
\caption{\label{fig:TestC} Test C results. The torsion balance pointed to the North. The magnet damper's orientation was flipped. There were 3 data points of each bar. Note that the scale of the Y axis is different from FIG.~\ref{fig:TestA}.}
\end{figure}

In Test C, we flipped the orientation of the magnetic damper, such that it was in the opposite orientation as in FIG.~\ref{fig:circuitNew}. Again we aligned the torsion balance to the North. We changed the return path 3 connectivity by altering the electrical connectivity at point C, 6 times in total, 3 times being connected and 3 times being disconnected, respectively. We grouped the measured rotary displacements according to the return path 3 connectivity. A two sided t-test gave a p-value of $0.0080$. The test results are illustrated in FIG.~\ref{fig:TestC}. 

At one time during Test C, we made a measurement with a steel washer placed on the beam near point A. The recorded rotary displacement was similar to those without the washer, suggesting that ferromagnetic material was unlikely to impair an experiment.

We also tested with different RF cable shapes and with different return path resistances (results observed but not recorded). Again the results suggested that the change of currents in the return paths and/or the change of shapes of the return paths could potentially cause the torsion balance's rotary displacement to change.

\section{\label{sec:Simple}The Simplified Experiment Apparatus\cite{SM,SMzip}}

\begin{figure}
\includegraphics[scale=0.83]{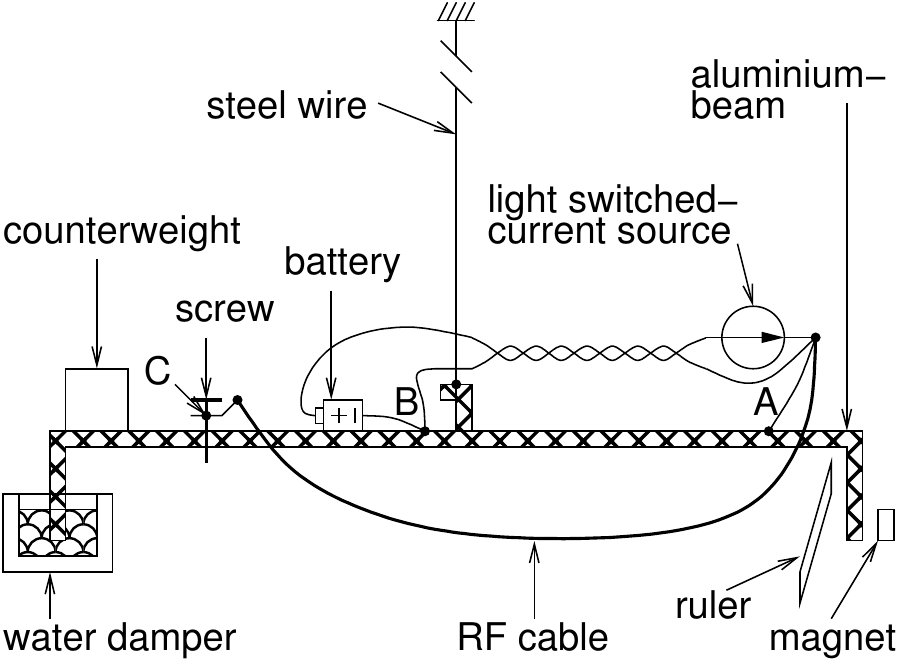}
\caption{\label{fig:circuitSimple} The simplified  experiment apparatus.}
\end{figure}

We simplify the experiment apparatus as illustrated in FIG.~\ref{fig:circuitSimple}. We relax the requirement of non-ferromagnetic material, the control of thermal disturbance, and the precise measurement provided by laser and mirror, while the experiment is still able to show the phenomenons we have observed. The Neodymium magnet can be replaced with a ceramic magnet. The  laser pointer used to switch the current source can be replaced with a bright torchlight. A video clip about the simplified experiment apparatus in testing is available at \url{https://www.youtube.com/watch?v=UsOee729YBM}.

\section{Discussion}
This experiment can be easily reproduced in a teaching laboratory. We believe it is useful for teaching the presented concepts to the students taking introductory physics courses. Its relationship with two recent experiments may help draw interest and dynamics to the classroom.

The first recent experiment was carried out by the Eagleworks team of NASA, attempting to detect thrusts from microwave driven resonance cavities \cite{Brady2014}. Their experiment was similar to ours, with the difference in that they used asymmetrical resonance cavities (the ``tapered cavity" or ``EmDrive", and the ``Cannae drive") while we use a symmetrical one (the copper cylinder with end caps), and in that they generated and fed microwave into the resonance cavities while we do not. They first measured a torsion balance's rotary displacement when microwave was fed into the resonance cavities, which were grounded. They then measured the torsion balance's rotary displacement when the resonance cavities were replaced with a 50 $\Omega$ resistive load, which was not likely grounded \cite{SM}. The RF cable shapes were also different in these two situations. The differences in measured torsion balance's rotary displacements were thought to be caused by thrusts generated by microwave bouncing back and forth inside the asymmetrical resonance cavities. Our experiment suggested that the differences could at least partially be explained by the alternations of grounding or alternations of the shapes of the return paths, or both, and the measured ``thrusts" were comparable in size with their results, putting their conclusion into question.

Another recent experiment was carried out by a research team at Dresden University of Technology \cite{Tajmar2015}. The researchers observed positive and negative rotary displacements of a torsion balance with opposite orientations of an asymmetrical microwave driven resonance cavity. We noticed that although they got rid of the magnetic damper, the Earth's magnetic field was not addressed. Also the magnets of the magnetron might have interacted with the power supply current. The researchers may need to consider these effects to reach meaningful conclusions.

\appendix

\section{\label{background}Background}
The Eagleworks team of NASA carried out an experiment (We call it the Brady experiment. We call their published paper the Brady paper.) to detect thrusts from microwave driven asymmetrical resonance cavities, including the tapered cavity (called ``EmDrive" elsewhere) and the Cannae Drive. They reported that thrusts were detected.

However, we believe that there is an alternative explanation to their findings. In the following we summarize some observations of their experiment that we think are important, upon which our alternative explanation is based.

In the Brady experiment, the authors used a torsion balance with magnetic damper to detect thrusts. The magnetic damper had strong grade N42 Neodymium magnets. Devices used to generate, amplify, and distribute microwave were mounted on, and grounded to, the torsion balance (see Figure 10 of the Brady paper for an example configuration). The power supply to the microwave amplifier was grounded to the supportive structure, which was electronically connected to the torsion balance (Figure 10 of the Brady paper). The RF cable that fed microwave into the resonance cavities or the resistive load did not have the same shape in different tests.

In the Brady experiment, there were null tests, with which microwave was fed into a 50 $\Omega$ resistive load, as well as resonance cavity tests, with which microwave was fed into the resonance cavities and let bouncing back and forth inside the resonance cavities. They recognized the existence of the Lorentz force caused by power supply current and the magnetic field from the magnetic damper. They used the null tests to remove the influence of the Lorentz force (section IV, subsection B of the Brady paper; also Figure 20 of the Brady paper).

Although according to Figure 10 of the Brady paper, the on and off of the microwave was controlled by the oscillator (VCO), Figure 20 of the paper showed that, during the null test, the amplifier power supply was turned on and off as well, where the 5.6 Ampere amplifier power supply current caused an average net thrust of 9.6 $\mu N$ when turned on. Furthermore, It was evident in Figure 19 of the paper, that during the tapered cavity test, the 5.6 Ampere power supply current was also turned on and off with the microwave, because the 9.6 $\mu N$ was subtracted from the raw thrust detected to yield net thrust, to remove the influence of the Lorentz force. This observation was consistent with the observation that cooling fan was not used to cool the amplifier (Figure 17, panel 3 of the Brady paper, compared to the pictures of the microwave amplifiers ZHL-100W-13+ or ZHL-30W-252+, which were available on their data sheets downloadable from the Mini-circuits website) thus there would have been heat dissipation problem if the amplifier were left on all the time. After all, the cooling fan could not be used at the first place, otherwise air flow disturbance would have been out of control.

Figure 17 of the Brady paper showed that the shell of the tapered cavity was grounded to the torsion balance through mounting screws and nuts when under testing. Figure 20 of the paper showed that the resistive load was placed on the torsion balance and was fixed to it with a plastic fastener. It was unlikely that the resistive load was grounded to the torsion balance, because its black appearance indicated that its enclosure (served as heat sink) was either painted or anodized.

We can summarize our observations of the Brady experiment as follows,
\begin{itemize}
\item There was a magnetic damper capable of generating magnetic field around it.
\item The shape of the RF cable that connected the microwave distributor and the resonance cavity or the resistive load was not the same in the different tests.
\item The power supply, the microwave amplifier/distributor and the microwave distributor were grounded.
\item The resonance cavities were grounded when under testing.
\item The resistive load was likely not grounded when under testing.
\item The 5.6 Ampere amplifier power supply current was turned on and off with the microwave.
\end{itemize}

Assuming these observations are true, we design our experiment to show that there is an alternative explanation to the thrusts they detected. Our experiment suggests that the influences of the Lorentz forces can not be removed simply by subtracting the thrust measured during the null test from that measured during the resonance cavity test. Thus it is questionable to attribute the detected net thrust to microwave bouncing inside the resonance cavities.

\section{Notes}

Appendix \ref{background} is taken from the supplemental material.  Other portions of the supplemental material are not made available to the arXiv version due to the size of the photographs. 

\nocite{*}
\bibliography{manuscript}
\end{document}